\documentclass[aps,pre,twocolumn,nofootinbib,floatfix,superscriptaddress,showpacs]{revtex4}
\usepackage{amsmath}
\usepackage{epsfig}

\input{tcilatex}

\begin{document}

\title{Proposal for detection of non-Markovian decay via current noise}
\author{Yueh-Nan Chen}
\email{yuehnan@mail.ncku.edu.tw}
\affiliation{Department of Physics and National Center for Theoretical Sciences, National
Cheng-Kung University, Tainan 701, Taiwan}
\author{Guang-Yin Chen}
\affiliation{Institute of Physics, National Chiao-Tung University, Hsinchu 300, Taiwan}

\begin{abstract}
We propose to detect non-Markovian decay of an exciton qubit coupled to
multi-mode bosonic reservoir via shot-noise measurements. Non-equilibrium
current noise is calculated for a quantum dot embedded inside a \textit{p-i-n%
} junction. An additional term from non-Markovian effect is obtained in the
derivation of noise spectrum. As examples, two practical photonic
reservoirs, photon vacuum with the inclusion of cut-off frequency and
surface plasmons, are given to show that the noise may become
super-Poissonian due to this non-Markovian effect. Utilizing the property of
super-radiance is further suggested to enhance the noise value.
\end{abstract}

\maketitle

\address{$^{1}$Department of Electrophysics, National Chiao Tung University,
Hsinchu 30050, Taiwan\\
$^{2}$Department of Physics, UMIST, P.O. Box 88, Manchester, M60 1QD, U.K.}

\address{$^{1}$Department of Electrophysics, National Chiao Tung University,
Hsinchu 30050, Taiwan} 
\address{$^{2}$Department of Physics, UMIST, P.O. Box 88, Manchester, M60
1QD, U.K.}

\address{Department of Electrophysics, National Chiao Tung University,
Hsinchu 300, Taiwan}





\textit{Introduction}.--Due to rapid progress of quantum information
science, great attention has been focused on the dynamics of systems
interacting with their surroundings. Radiative decay of a two-level atom
maybe one of the most obvious examples in this issue and can be traced back
to such early works as that of Albert Einstein in 1917. \cite{1} While
Markovian approximation is widely adopted to treat decoherence and
relaxation problems, non-Markovian dynamics of qubit systems have attracted
increasing attention lately. \cite{2}

Turning to solid state systems, an exciton in a quantum dot (QD) can be
viewed as a two-level system. Radiative properties of QD excitons, such as
super-radiance \cite{3} and Purcell effect \cite{4}, have attracted great
attention during the past two decades. Utilizing QD excitons for quantum
gate operations have also been demonstrated experimentally. \cite{5} With
the advances of fabrication technologies, it is now possible to embed QDs
inside a \textit{p-i-n} structure \cite{6}, such that the electron and hole
can be injected separately from opposite sides. This allows one to examine
the exciton dynamics in a QD via electrical currents \cite{7}.

Recently, the interest in measurements of shot noise in quantum transport
has risen owing to the possibility of extracting valuable information not
available in conventional dc transport experiments \cite{8}. In this work,
we propose to detect non-Markovian decay of a exciton qubit via the current
noise of a QD \textit{p-i-n} junction. Without making Markovian
approximation to the exciton-boson interaction, we analytically show that
the Fano factor (zero-frequency noise) may become super-Poissonian. As
examples, two practical photonic environments, photon vacuum with the
inclusion of cut-off frequency and surface plasmons, are given to show this
non-Markovian effect. To enhance the noise value, we further suggest
utilizing the property of super-radiance. 
\begin{figure}[th]
\includegraphics[width=7cm]{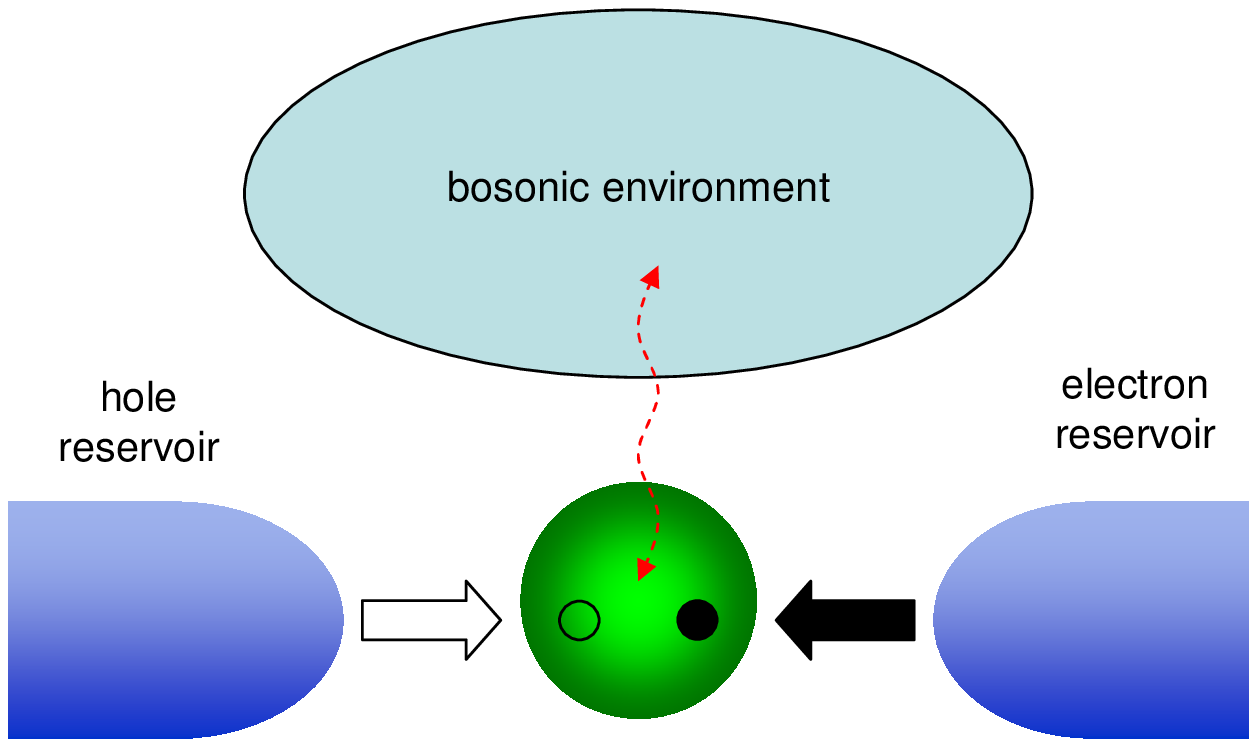}
\caption{{}(Color online) Schematic view of a QD \textit{p-i-n} junction
with its exciton coupled to a bosonic environment.}
\end{figure}

\textit{The model.--}QDs can now be embedded in a \textit{p-i-n} junction,
such that many applications can be accomplished by electrical control. As
shown in Fig. 1, we wish to see non-Markovian effect between the system and
reservoir via measurements of electrical currents. For simplicity, both the
hole and electron reservoirs of the \textit{p-i-n} junction are assumed to
be in thermal equilibrium. For the physical phenomena we are interested in,
the Fermi level of the \textit{p(n)}-side hole (electron) is slightly lower
(higher) than the hole (electron) subband in the dot. After a hole is
injected into the hole subband in the QD, the \textit{n}-side electron can
tunnel into the exciton level because of the Coulomb interaction between the
electron and hole. Thus, we may introduce the three dot states: $\left|
0\right\rangle =\left| 0,h\right\rangle $, $\left| \uparrow \right\rangle
=\left| e,h\right\rangle $, and $\left| \downarrow \right\rangle =\left|
0,0\right\rangle $, where $\left| 0,h\right\rangle $ means there is one hole
in the QD,\ $\left| e,h\right\rangle $ is the exciton state, and $\left|
0,0\right\rangle $ represents the ground state with no hole and electron in
the QD. One might argue that one can not neglect the state $\left|
e,0\right\rangle $ for real devices since the tunable variable is the
applied voltage. This can be resolved by fabricating a thicker barrier on
the electron side so that there is little chance for an electron to tunnel
in advance \cite{7}. Thus, the couplings of the dot states to the electron
and hole reservoirs are to be described by the standard tunnel Hamiltonian

\begin{equation}
H_{T}=\sum_{\mathbf{q}}(V_{\mathbf{q}}c_{\mathbf{q}}^{\dagger }\left|
0\right\rangle \left\langle \uparrow \right| +W_{\mathbf{q}}d_{\mathbf{q}%
}^{\dagger }\left| 0\right\rangle \left\langle \downarrow \right| +H.c.),
\end{equation}%
where $c_{\mathbf{q}}$ and $d_{\mathbf{q}}$ are the electron operators in
the right and left reservoirs, respectively. $V_{\mathbf{q}}$ and $W_{%
\mathbf{q}}$ couple the channels $\mathbf{q}$ of the electron and the hole
reservoirs. The interaction between the exciton qubit and its bosonic
environment is written as

\begin{eqnarray}
H_{ex-bosonic} &=&\sum_{k}D_{k}b_{k}^{\dagger }\left| \downarrow
\right\rangle \left\langle \uparrow \right| +H.c.  \notag \\
&=&\left| \downarrow \right\rangle \left\langle \uparrow \right| X+\left|
\uparrow \right\rangle \left\langle \downarrow \right| X^{\dagger },
\end{eqnarray}%
where $X=\sum_{k}D_{k}b_{k}^{\dagger }$, $b_{k}^{\dagger }$ denotes the
creation operator of the bosonic reservoir, and $D_{k}$ describes the
system-reservoir coupling.

With Eqs. (1) and (2), one can now write down the equation of motion for the
reduced density operator

\begin{eqnarray}
\frac{d}{dt}\rho (t) &=&-Tr_{res}\int_{0}^{t}dt^{\prime
}[H_{T}(t)+H_{ex-bosonic}(t),  \notag \\
&&[H_{T}(t^{\prime })+H_{ex-bosonic}(t^{\prime }),\widetilde{\Xi }(t^{\prime
})]],
\end{eqnarray}%
where $\widetilde{\Xi }(t^{\prime })$ is the total density operator. Note
that the trace in Eq. (3) is taken with respect to both bosonic and
electronic reservoirs. If the couplings to the electron and hole reservoirs
are weak, it is reasonable to assume that the standard Born-Markov
approximation with respect to the electronic couplings is valid. In this
case, multiplying Eq. (3) by $\overset{\wedge }{n_{\uparrow }}=\left|
\uparrow \right\rangle \left\langle \uparrow \right| $ and $\overset{\wedge }%
{n_{\downarrow }}=\left| \downarrow \right\rangle \left\langle \downarrow
\right| $, the equations of motions can be written as

\begin{eqnarray}
\frac{\partial }{\partial t}\binom{\overset{\wedge }{\left\langle
n_{\uparrow }\right\rangle }_{t}}{\overset{\wedge }{\left\langle
n_{\downarrow }\right\rangle }_{t}} &=&\int dt^{\prime }\binom{%
-A(t-t^{\prime })\overset{\wedge }{\left\langle n_{\uparrow }\right\rangle }%
_{t^{\prime }}}{A(t-t^{\prime })\overset{\wedge }{\left\langle n_{\downarrow
}\right\rangle }_{t^{\prime }}}  \notag \\
&&+\left[ 
\begin{array}{cc}
-\Gamma _{L} & -\Gamma _{L} \\ 
0 & -\Gamma _{R}%
\end{array}%
\right] \binom{\overset{\wedge }{\left\langle n_{\uparrow }\right\rangle }%
_{t}}{\overset{\wedge }{\left\langle n_{\downarrow }\right\rangle }_{t}}+%
\binom{\Gamma _{L}}{0},
\end{eqnarray}%
where $\Gamma _{L}$ $=2\pi \sum_{\mathbf{q}}V_{\mathbf{q}}^{2}\delta
(\varepsilon _{\uparrow }-\varepsilon _{\mathbf{q}}^{\uparrow })$ , $\Gamma
_{R}=2\pi \sum_{\mathbf{q}}W_{\mathbf{q}}^{2}\delta (\varepsilon
_{\downarrow }-\varepsilon _{\mathbf{q}}^{\downarrow })$, and $\varepsilon
=\hbar \omega _{0}=\varepsilon _{\uparrow }-\varepsilon _{\downarrow }$ is
the energy gap of the QD exciton. Here, $A(t-t^{\prime })\equiv
C(t-t^{\prime })+C^{\ast }(t-t^{\prime })$ can be viewed as the (bosonic)
reservoir correlation function with the function $C$ \ defined as $%
C(t-t^{\prime })$ $\equiv \left\langle X_{t}X_{t^{\prime }}^{\dagger
}\right\rangle _{0}$. The appearance of the two-time correlation is
attributed to that in the derivation of Eq. (4), we only assume the Born
approximation to the bosonic reservoir, i.e. the Markovian one is not made.

One can now define the Laplace transformation for real $z,$

\begin{eqnarray}
C_{\varepsilon }(z) &\equiv &\int_{0}^{\infty }dte^{-zt}e^{i\varepsilon
t}C(t)  \notag \\
n_{\uparrow }(z) &\equiv &\int_{0}^{\infty }dte^{-zt}\overset{\wedge }{%
\left\langle n_{\uparrow }\right\rangle }_{t}\text{ \ }etc.,\text{ }z>0
\end{eqnarray}%
and transform the whole equations of motion into $z$-space,

\begin{eqnarray}
n_{\uparrow }(z) &=&-A_{\varepsilon }(z)n_{\uparrow }(z)/z+\frac{\Gamma _{L}%
}{z}[1/z-n_{\uparrow }(z)-n_{\downarrow }(z)],  \notag \\
n_{\downarrow }(z) &=&A_{\varepsilon }(z)n_{\downarrow }(z)/z-\frac{\Gamma
_{R}}{z}n_{\downarrow }(z).
\end{eqnarray}%
These equations can then be solved algebraically, and the tunnel current
from the hole--side barrier, $\overset{\wedge }{I}_{R}=-e\Gamma _{R}\overset{%
\wedge }{\left\langle n_{\downarrow }\right\rangle }_{t}$, can in principle
be obtained by performing the inverse Laplace transformation. Depending on
the complexity of the correlation function $C(t-t^{\prime })$ in the time
domain, this can be a formidable task which can however be avoided if one
directly seeks the quantum noise.

In a quantum conductor in nonequilibrium, electronic current noise
originates from the dynamical fluctuations of the current around its
average. To study correlations between carriers, we relate the exciton
dynamics with the hole reservoir operators by introducing the degree of
freedom $n$ as the number of holes that have tunneled through the hole-side
barrier, and write 
\begin{gather}
\overset{\cdot }{n}_{0}^{(n)}(t)=-\Gamma _{L}n_{0}^{(n)}(t)+\Gamma
_{R}n_{\downarrow }^{(n-1)}(t),  \notag \\
\overset{\cdot }{n}_{\uparrow }^{(n)}(t)+\overset{\cdot }{n}_{\downarrow
}^{(n)}(t)=(\Gamma _{L}-\Gamma _{R})n_{0}^{(n)}(t).
\end{gather}%
Eqs. (7) allow us to calculate the particle current and the noise spectrum
from $P_{n}(t)=n_{0}^{(n)}(t)+n_{\uparrow }^{(n)}(t)+n_{\downarrow
}^{(n)}(t) $ which gives the total probability of finding $n$ electrons in
the collector by time $t$. In particular, the noise spectrum $S_{I_{R}}$ can
be calculated via the MacDonald formula \cite{9,10}, 
\begin{equation}
S_{I_{R}}(\omega )=2\omega e^{2}\int_{0}^{\infty }dt\sin (\omega t)\frac{d}{%
dt}[\left\langle n^{2}(t)\right\rangle -(t\left\langle I\right\rangle )^{2}],
\end{equation}%
where $\frac{d}{dt}\left\langle n^{2}(t)\right\rangle =\sum_{n}n^{2}\overset{%
\cdot }{P_{n}}(t)$. With the help of counting statistics \cite{10}, one can
obtain 
\begin{gather}
S_{I_{R}}(\omega )=2eI\{1+ \\
\Gamma _{R}[\frac{A(i\omega )\Gamma _{L}}{-A(i\omega )\Gamma _{L}\Gamma
_{R}+(A(i\omega )+i\omega )(\Gamma _{L}+i\omega )(\Gamma _{R}+i\omega )}+ 
\notag \\
\frac{A(-i\omega )\Gamma _{L}}{-A(-i\omega )\Gamma _{L}\Gamma
_{R}+(A(-i\omega )-i\omega )(\Gamma _{L}-i\omega )(\Gamma _{R}-i\omega )}]\},
\notag
\end{gather}%
where $A(z)\equiv C_{\varepsilon }(z)+C_{\varepsilon }^{\ast }(z)$.

As can be seen from Eq. (9), the noise spectrum indeed contains the
information of memory effect, i.e. $A(i\omega )$ and $A(-i\omega )$.
However, it is not easy to see how it affects the noise. We thus take the
zero-frequency limit ($\omega \rightarrow 0$), and an analytical solution
with physical meaning is obtained:

\begin{align}
F& =S_{I_{R}}(\omega =0)/2e\left\langle I\right\rangle  \notag \\
& =1-\frac{2\Gamma _{L}\Gamma _{R}[Re[A(0)]\Gamma
_{L}+Re[A(0)](Re[A(0)]+\Gamma _{R})]}{[Re[A(0)]\Gamma _{R}+\Gamma
_{L}(Re[A(0)]+\Gamma _{R})]^{2}}  \notag \\
& +\frac{2Im[\frac{\partial A(\emph{iw})}{\partial \emph{w}}|_{\emph{w}%
=0}]\Gamma _{L}^{2}\Gamma _{R}^{2}}{[Re[A(0)]\Gamma _{R}+\Gamma
_{L}(Re[A(0)]+\Gamma _{R})]^{2}}.
\end{align}%
If one makes Markovian approximation to the bosonic reservoir, the third
term in Eq. (10) vanishes and the Fano factor ($F$) is further reduced to
usual sub-Poissonian result. The question here is that whether this
additional term is positive or not, such that the noise feature may become
super-Poissonian. To answer this, let us now consider real bosonic
environments.

\textit{Surface plasmons}.--The collective motions of an electron gas in a
metal or semiconductor are known as the plasma oscillations. The
non-vanishing divergence of the electric field $\overrightarrow{E}$, $\nabla
\cdot \overrightarrow{E}\neq 0$, in the bulk material gives rise to the well
known bulk plasma modes, characterized by the plasma frequency $\omega
_{p}=(4\pi n_{0}e^{2}/m)^{1/2}$, where $m$ and $e$ are the electronic mass
and charge and $n_{0}$ is the electron density. In the presence of surface,
however, the situation becomes more complicated. Not only the bulk modes are
modified, but also the surface modes can be created. \cite{11} Like the bulk
modes, surface plasmons can be excited by incident electrons or photons. %
\cite{12} Many works were devoted to the study of radiative decay into
surface plasmons. \cite{13} Recently, it is now possible to fabricate QDs
evanescently coupled to surface plasmons, such that enhanced fluorescence
are observed. \cite{14} Based on these new developments, it is plausible to
assume that the QD \textit{p-i-n} junction is close to a metal surface. This
allows one to examine non-Markovian effect from surface plasmons.

\begin{figure}[th]
\includegraphics[width=8cm]{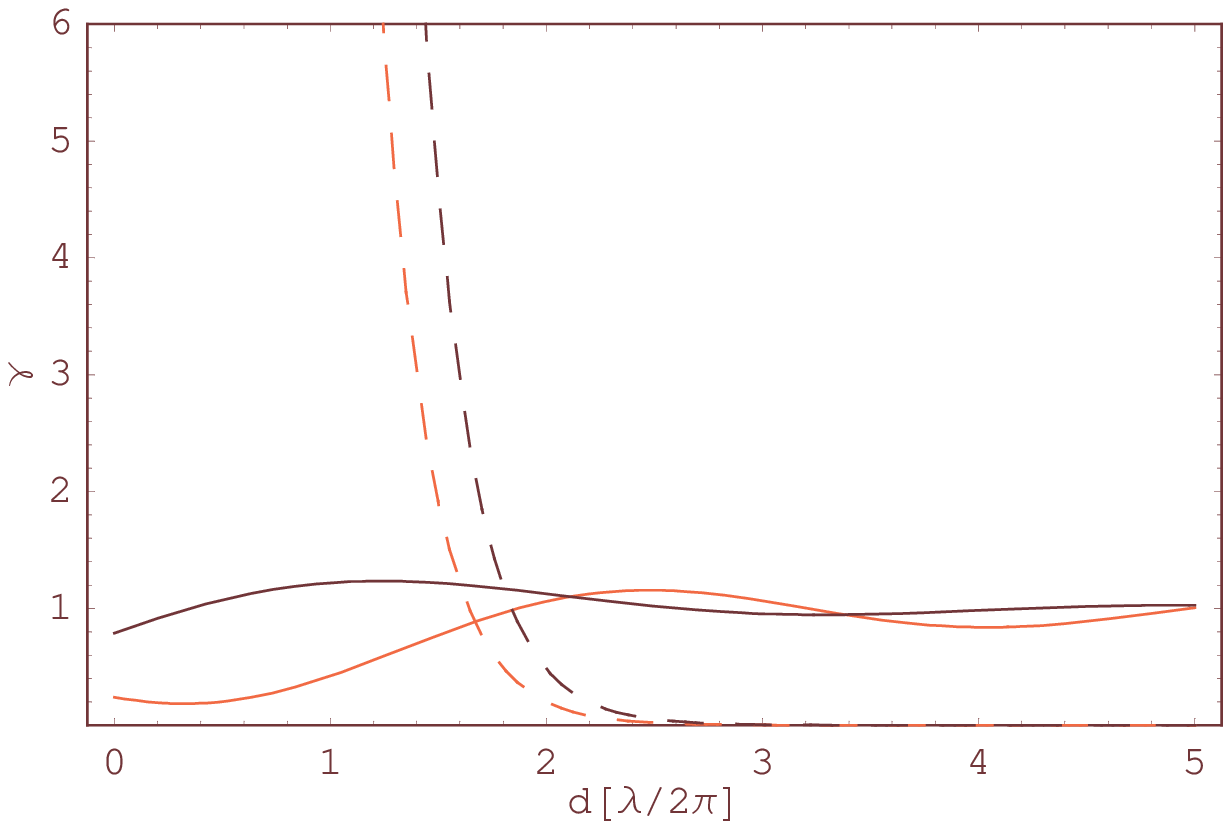}
\caption{{}(Color online) Radiative decay rate of QD exciton in front of a
silver surface with distance $d$ (in units of $\protect\lambda /2\protect\pi 
$, where $\protect\lambda $ is the wavelength of the emitted photon). The
plasma oscillation energy $\hbar \protect\omega _{p}$ of silver and exciton
bandgap energy $\hbar \protect\omega _{0}$ are $3.76eV$ and $1.39eV$ %
\protect\cite{6}, respectively. The black dashed (solid) line represents the
decay into the surface plasmons (photons) as the exciton dipole moment $%
\protect\overset{\wedge }{p}$ is oriented perpendicular to the surface. The
red lines are the case for $\protect\overset{\wedge }{p}$ parallel to the
surface.}
\end{figure}

When a semiconductor QD is near a metal surface, the vector potential to the
QD exciton can be decomposed into contributions from s- and p-polarized
photons and surface plasmons \cite{15}: 
\begin{equation}
A(\overrightarrow{r},t)=A^{s}(\overrightarrow{r},t)+A^{p}(\overrightarrow{r}%
,t)+A_{sp}(\overrightarrow{r},t).
\end{equation}%
Fig. 2 shows the corresponding radiative decay rates of a QD exciton in
front of a silver surface. It is evident that at short distances radiative
decay is dominated by surface plasmons. Since we are interested in the
effect from surface plasmons, we thus keep QD in this regime, and consider
only the interaction from surface plasmons:

\begin{eqnarray}
H_{ex-sp} &=&\sum_{\mathbf{k}}(\frac{4\pi \omega _{k}^{2}}{\hbar Acp_{k}}%
)^{1/2}[\left| \uparrow \right\rangle \left\langle \downarrow \right| (%
\overset{\wedge }{k}\cdot \overset{\wedge }{p}+i\frac{k}{\nu _{0}}\cdot 
\overset{\wedge }{p})a_{k}e^{i\mathbf{k\cdot \rho }-\nu _{0}z}  \notag \\
&&+H.c.].
\end{eqnarray}%
Here, we have chosen cylindrical coordinates $\overrightarrow{r}=(%
\overrightarrow{\rho },z)$ in the half-space $z\geq 0$; $\overrightarrow{k}$
is a two dimensional wave vector in the metal surface of area $A$. $a_{k}$
is the annihilation operator of surface plasmon and $\overset{\wedge }{p}$
is the transition dipole moment. The surface-plasmon frequency $\omega _{k}$
and the parameters, $\nu _{0}$ and $p_{k}$, are given by%
\begin{eqnarray}
\omega _{k}^{2} &=&\frac{1}{2}\omega _{p}^{2}+ck^{2}-(\frac{1}{4}\omega
_{p}^{4}+c^{2}k^{4})^{1/2};\nu _{0}=k^{2}-\omega _{k}^{2}/c^{2};  \notag \\
p_{k} &=&\frac{\epsilon ^{4}(\omega _{k})-1}{[-\epsilon (\omega
_{0})-1]^{1/2}}\frac{1}{\epsilon ^{2}(\omega _{k})},
\end{eqnarray}%
where $\epsilon (\omega _{k})=1-$ $\omega _{p}^{2}/\omega _{k}^{2}$ is the
dielectric function of the metal. By replacing $H_{ex-bosonic}$ with $%
H_{ex-sp}$, one can go through the procedure to obtain the current noise.

\begin{figure}[th]
\includegraphics[width=8cm]{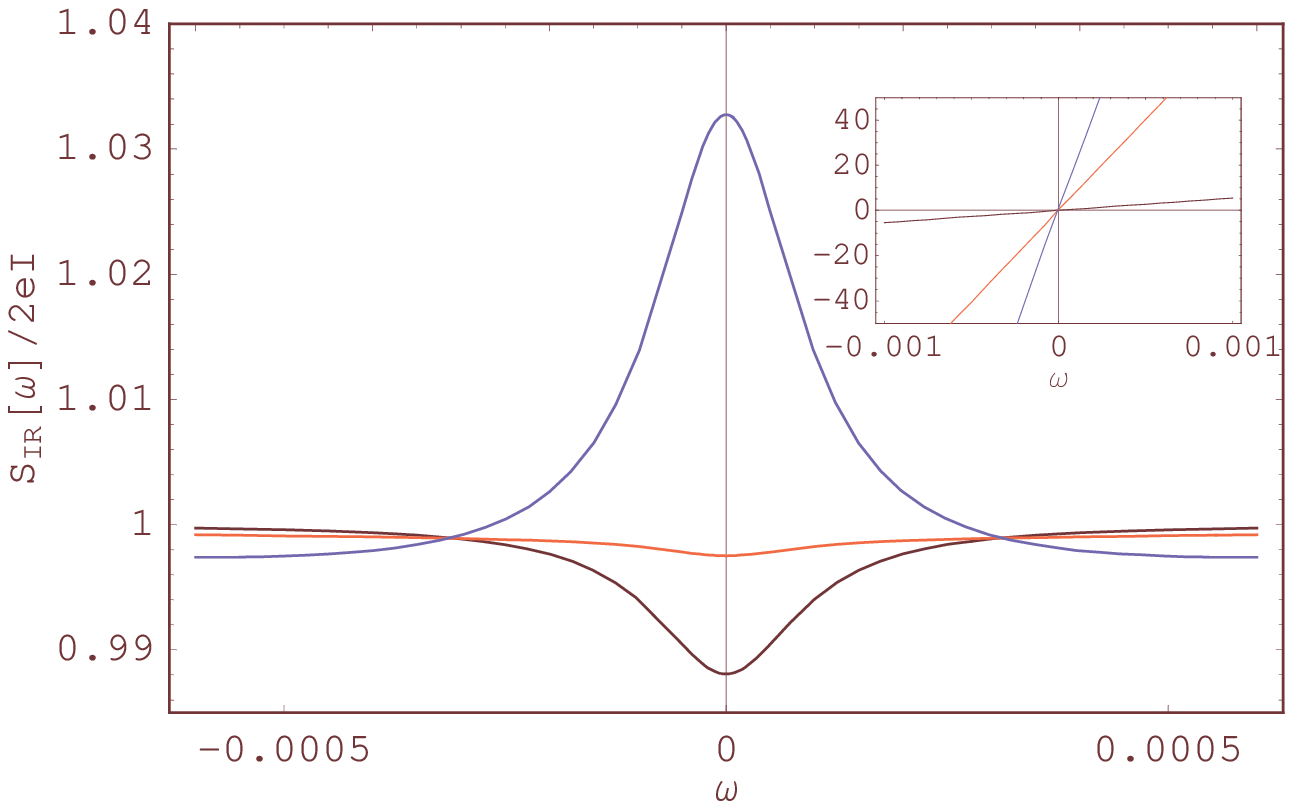}
\caption{{}(Color online) Shot-noise spectrums of QD excitons in front of a
silver surface. \ The black, red, and blue lines represent the results of
various dot-surface distances: $d=0.1,$ $0.045$, and $0.03$ (in units of $%
\protect\lambda /2\protect\pi \approx 1423\protect\overset{\circ }{A})$,
respectively. The inset shows the corresponding curves of the imaginary part
of $A(i\protect\omega )$.}
\end{figure}

The shot-noise spectrum of InAs QD excitons is numerically displayed in Fig.
3, where the tunneling rates, $\Gamma _{L}$ and $\Gamma _{R}$, are assumed
to be equal to $10^{-4}\omega _{0}$ and $10^{-3}\omega _{0}$, respectively.
The plasma oscillation energy $\hbar \omega _{p}$ of silver and exciton
bandgap energy $\hbar \omega _{0}$ are $3.76eV$ and $1.39eV$. One knows from
Fig. 2 that there is no essential difference in physics for different
orientations of the exciton dipole moment. Therefore, in plotting the figure
the dipole moment $\overset{\wedge }{p}$ is assumed to be along $\overset{%
\wedge }{z}$ direction for simplicity. Without making Markovian
approximation, the black, red, and blue lines represent the results for
different dot-surface distances: $d=0.1$, $0.045$, and $0.03$ (in unit of $%
\lambda /2\pi \approx 1423\overset{\circ }{A}$), respectively. As seen, the
Fano factor gradually changes from sub-Poissonian noise to super-Poissonian
one as the QD is moving toward the surface. This proves that the additional
term in Eq. (10) can change the noise feature. The inset of Fig. 3
numerically shows the imaginary part of $A(i\omega )$. As the QD is closer
to the silver surface, the slope becomes steeper, which coincides with the
analytical result of Eq. (10).

Reasons for super-Poissonian noise actually depend on the details of the
device structures. For examples, positive correlations due to resonant
tunneling states \cite{16}, noise enhancement due to quantum entanglement %
\cite{17}, spin-flip co-tunneling processes \cite{18}, non-Markovian
coupling between dot and leads \cite{19}, and quantum shuttle effect \cite%
{20}. The underlying physical picture in our case maybe similar to a recent
work by Djuric \textit{et al}. \cite{21} They considered the tunneling
problem through a QD connected coherently to a nearby single-level dot,
which is not connected to the left and right leads. In this case, the
coherent hopping to the nearby dot also gives an extra ''positive'' term to
the Fano factor. The explanation is that the coming electron can either
tunnel out of the original dot directly, or travel to the nearby dot and
come back again. This indirect path is the origin of the super-Poissonian
noise. In our case, as the exciton decays into surface plasmon, the
non-Markovian effect from the plasmon reservoir may re-excite it now and
then, such that the Fano factor is enhanced. One notes that this kind of
enhancement due to quantum coherence has recently been observed in the
tunneling through a stack of coupled quantum dots \cite{22} and explained
theoretically \cite{23}. 
\begin{figure}[th]
\includegraphics[width=8cm]{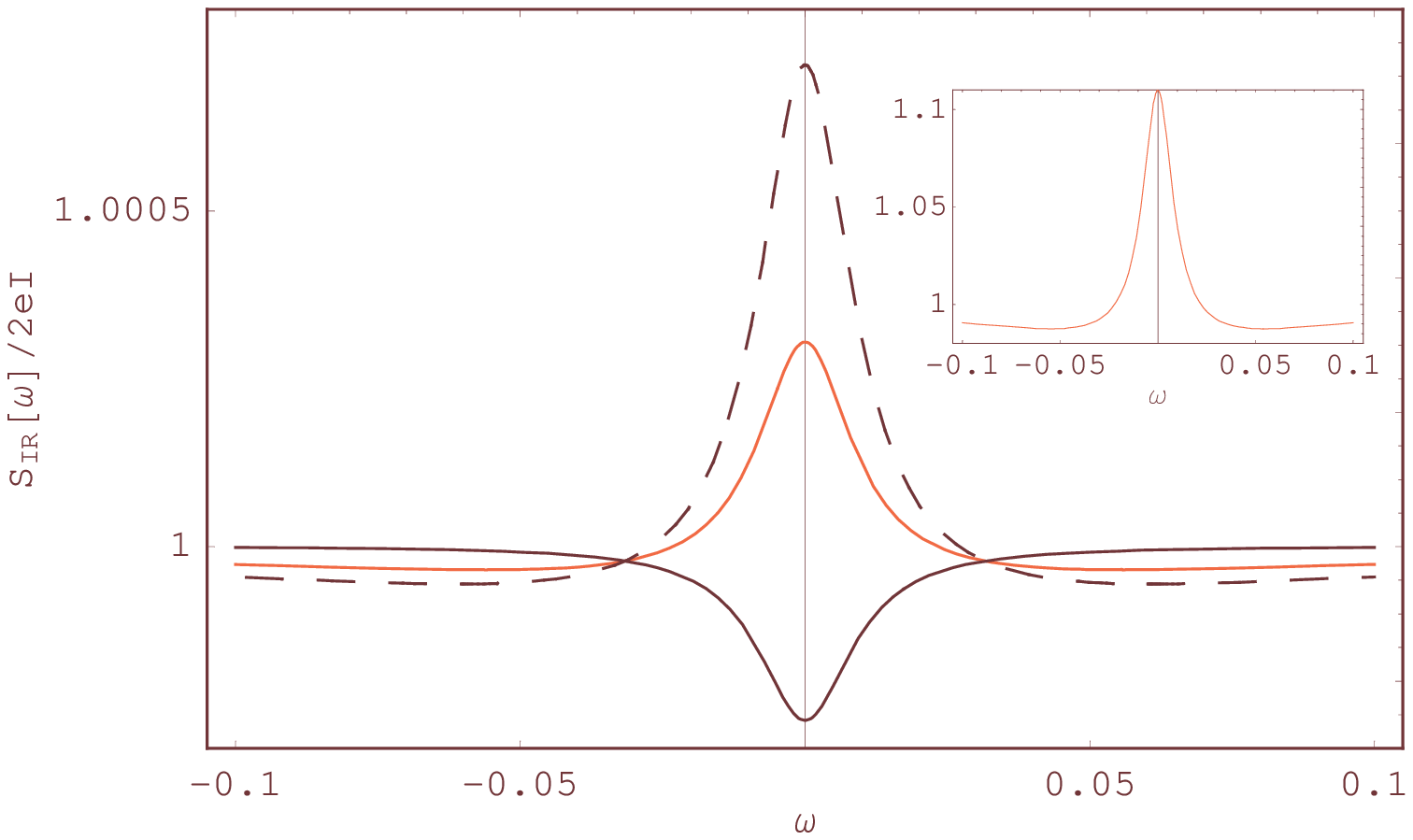}
\caption{{}(Color online) Shot-noise spectrums of QD excitons in the
presence of Lorentzian cutoff. \ Sub-Poissonian noise represented by the
black line is the result of Markovian approximation. Super-Poissonian noise
(red and dashed lines) is the consequence of non-Markovian effect. To plot
the figure, the exciton spontaneous lifetime ($=1/\protect\gamma $) used
here is $1.3ns$, and the cutoff frequency for red (dashed) line is $9\times
10^{16}$ $(1.2\times 10^{17})Hz$. Inset: Noise increased by the enhancemet
of the effective dipole moment (400 times) via super-radiance. }
\end{figure}

\textit{The cutoff frequency}.--To see whether this non-Markovian effect is
general, let us return to the old quantum electrodynamic (QED) problem:
spontaneous emission. Under Markovian approximation, the emission rate of a
two-level atom in free space can be easily obtained via the Fermi's Golden
rule and is give by $\gamma =2\pi \sum_{\mathbf{q}}$ $\left| D_{\mathbf{q}%
}\right| ^{2}\delta (\omega _{0}-c\left| \mathbf{q}\right| )$, where $D_{%
\mathbf{q}}$ is the atom-reservoir coupling strength. It's frequency
counterpart is written as $\Delta \omega =$ $\mathcal{P}\int d\mathbf{q}$ $%
\left| D_{\mathbf{q}}\right| ^{2}/(\omega _{0}-c\left| \mathbf{q}\right| ),$
where $\mathcal{P}$ denotes the principal integral. To remove the divergent
problem from the integration, one can, for example, include the concept of
cutoff frequency to renormalize the frequency shift. \cite{24} In this case,
the exciton-photon coupling is described by the Hamiltonian:

\begin{equation}
H_{ex-Ph}=\sum_{k}\frac{1}{(1+(\omega _{k}/\omega _{B})^{2})^{2}}%
D_{k}b_{k}^{\dagger }\left| \downarrow \right\rangle \left\langle \uparrow
\right| +H.c.,
\end{equation}%
where the introduced Lorentzian cutoff contains the nonrelativistic cutoff
frequency $\omega _{B}\approx c/a_{B}$, with $a_{B}$ being the effective
Bohr radius of the exciton. \cite{25} Replacing $H_{ex-bosonic}$ by $%
H_{ex-Ph}$, one can obtain the corresponding noise spectrum
straightforwardly. As shown in Fig. 4, the Fano factor is sub-Poissonian
(black line) under Markovian approximation, while it may become
super-Poissonian (as shown by the dashed and red lines) with the
consideration of non-Markovian effect from the Lorentzian cutoff.

One also finds that the magnitude of the Fano factor depends on the cutoff
frequency $\omega _{B}$. With the increasing of $\omega _{B}$, the Fano
factor becomes lager (the dashed line). This phenomenon allows one to
examine the cutoff frequency in QED. However, one might argue that the value
of the super-Poissonian noise is extremely small and may not be observable
in real experiments. To overcome this obstacle, we suggest making use of the
property of collective decay (super-radiance). \cite{26} For example, one
can, instead of the QD, insert a quantum well (QW) into the \textit{p-i-n}
junction. The interaction between the (two-dimensional) QW exciton and the
photon can be written as \cite{27}

\begin{eqnarray}
H^{\prime } &=&\sum_{\mathbf{q}nm}\sum_{k_{z}}\frac{1}{(1+(\omega
_{k}/\omega _{B})^{2})^{2}}\frac{e}{m_{0}c}\sqrt{\frac{2\pi \hbar c}{(%
\mathbf{q}^{2}+k_{z}^{2}\mathbf{)}^{1/2}v}}  \notag \\
&&\times (\mathbf{\epsilon }_{\mathbf{q}k_{z}}\cdot \mathbf{A}_{\mathbf{q}%
nm})b_{\mathbf{q}k_{z}}c_{\mathbf{q}nm}^{\dagger }+\mathbf{h.c.},
\end{eqnarray}%
where

\begin{equation}
\mathbf{A}_{\mathbf{q}nm}=\sqrt{N}\sum_{\mathbf{\rho }}F_{nm}^{\ast }(%
\mathbf{\rho })\int d^{2}\mathbf{\tau }w_{c}^{\ast }(\mathbf{\tau }-\mathbf{%
\rho })(-i\hbar \mathbf{\nabla })w_{v}(\mathbf{\tau })\text{.}
\end{equation}%
Here, $c_{\mathbf{q}}^{\dagger }$ and $b_{\mathbf{q}k_{z}}$ stand for the
exciton and photon operators. $\mathbf{\epsilon }_{\mathbf{q}k_{z}}$ is the
polarization vector of the photon. $F_{nm}(\mathbf{\rho })$ is the
two-dimensional hydrogenic wavefunction of the exciton with quantum number $%
n $ and $m$. $w_{c}(\mathbf{\tau })$ and $w_{v}(\mathbf{\tau })$ are,
respectively, the Wannier functions for the conduction band and the valence
band.

With the Hamiltonian in Eq. (15), the radiative decay rate of the QW exciton
can be obtained straightforwardly

\begin{equation}
\gamma _{\mathbf{q}nm}\sim \gamma _{0}(\lambda /d)^{2},
\end{equation}%
where $\gamma _{0}$ is the decay rate of a lone exciton, $\lambda $ is the
wavelength of the emitted photon, and $d$ is the lattice constant of the
material. The enhanced rate in Eq. (17) implies the coherent contributions
from the lattice atoms within half a wavelength or so. In another word, one
can say that the effective dipole moment of the QW exciton is enhanced by a
factor of $(\lambda /d)^{2}$ \cite{27}. From Eq. (10), Fig. 2, and inset of
Fig. 3, we know that an enhanced rate somehow implies a larger Fano factor.
Consider the real experimental values \cite{28}, the observed enhancement is
around several hundred times the lone exciton. We thus plot the Fano factor
in the inset of Fig. 4. As can be seen, the value of super-Poissonian noise
is greatly enhanced by super-radiance. This gives a better chance to observe
the mentioned effect. Another possible candidate for the enhancement is the
uniform QD-arrays. \cite{29} Within the collective decay area defined by $%
\lambda ^{2}$, the effective dipole moment may also be enhanced by a factor
of $(\lambda /r)^{2}$ , where $r$ is the dot-lattice constant.

Finally, we note that recent advances in fabrication nanotechnologies have
made it possible to grow high quality nanowires \cite{30}, in which cavity
QED phenomena can be revealed via surface plasmons. \cite{31} It is likely
that similar effects will appear if the QD \textit{p-i-n} junction is
coupled to the \textit{channel plasmons}. Even more, since the dispersion
relation in cylindrical interface is much more complex (for example, it
contains both real and virtual modes \cite{32}), the corresponding
shot-noise spectrums are expected to give more information about the
non-Markovian effect. Further investigations in this direction certainly put
such a system more useful in the fields of quantum transport and cavity QED.

We would like to thank Prof. T. Brandes and Prof. E. Sch\"{o}ll at
Technische Universit\"{a}t Berlin, Prof. J. Y. Hsu, and Prof. S. H. Chang at
Institute Electro-Optical Engineering for helpful discussions. This work is
supported partially by the National Science Council, Taiwan under the grant
number NSC 95-2112-M-006-031-MY3.

\end{document}